\documentclass[superscriptaddress, layout=twocolumn]{achemso}
\usepackage{graphicx}
\usepackage{amssymb}
\usepackage{epstopdf}
\usepackage{amsmath,dsfont, commath}
\usepackage{bm}
\usepackage{braket}
\usepackage{changes}
\usepackage{hyperref}
\usepackage{tensor}
\newcommand{\be}{\begin{equation}}
	\newcommand{\ee}{\end{equation}}
\newcommand{\bea}{\begin{eqnarray}}
	\newcommand{\eea}{\end{eqnarray}}

\newcommand{\ba}{\begin{array}}
	\newcommand{\ea}{\end{array}}

\newcommand{\se}{Schr\"{o}dinger equation}
\newcommand{\bl}{\begin{flalign}}
	\newcommand{\enl}{\end{flalign}}

\newcommand{\mc}[1]{\mathcal{#1}}

\newcommand{\tdse}{time-dependent Schr\"{o}dinger equation}

\usepackage{cleveref}

\newcommand{\half}{\frac{1}{2}}

\renewcommand{\bf}[1]{\mathbf{#1}}

\SectionNumbersOn

\usepackage{caption}

\crefname{equation}{eq}{eq}

\makeatletter
\renewcommand{\cref}[1]{eq~\ref{#1}}
\makeatother

	\title{Making peace with random phases: Ab initio conical intersection dynamics in random gauges} 
\author{Xiaotong Zhu}
    \author{Bing Gu}
\email{gubing@westlake.edu.cn}
\affiliation{Department of Chemistry and Department of Physics, Westlake University, Hangzhou, Zhejiang 310030, China}
\affiliation{Institute of Natural Sciences, Westlake Institute for Advanced Study, Hangzhou, Zhejiang 310024, China}

\begin{document}

\begin{abstract}
    Ab initio modeling of conical intersection dynamics is crucial for various photochemical, photophysical, and biological processes. 
    However, adiabatic electronic states obtained from electronic structure computations involve random phases, or more generally, random gauge fixings, which hampers the modeling of nonadiabatic molecular dynamics.  
    Here we develop a random-gauge local diabatic representation that allows an exact modeling of conical intersection dynamics directly using the adiabatic electronic states with phases randomly assigned during the electronic structure computations.  
    Its utility is demonstrated by an exact ab initio modeling of the two-dimensional Shin-Metiu model with and without an external magnetic field. 
    Our results provide a simple approach to integrating the electronic structure computations into non-adiabatic quantum dynamics, thus paving the way for ab initio modeling of conical intersection dynamics.   
\end{abstract}

\maketitle

\section{Introduction}

Conical intersections (CIs), ubiquitous in polyatomic molecules \cite{truhlar2003}, are the transition states for a wide range of ultrafast photoinduced processes, such as nonradiative electronic transitions, photophysical phenomena, and photoisomerization, photodissociation, light-matter interaction, and energy transfer \cite{ domcke2012,koppel2004, larson2020, levine2007, mead1979, mead1992, joubert-doriol2013, bersuker2006, gu2020b, gu2020c}. 
Around CIs, the Born-Oppenheimer approximation breaks down,  the electronic and nuclear motion becomes strong correlated. 
Even when the CI is energetically inaccessible, it can still impact the adiabatic nuclear dynamics through geometric phase effects \cite{xie2019}.  
Ab initio modeling of conical intersection dynamics, a quantum mechanical approach that directly calculates the electronic structure and nuclear quantum dynamics  in the presence of CIs without relying on empirical or fitted parameters, provides a detailed and accurate description of the non-adiabatic quantum molecular dynamics from first principles and sheds light on the influence of CIs on electron-nuclear correlation, reactivity, and spectroscopic observables.  


Ab initio modeling  employs electronic structure methods to build the potential energy surfaces required for nuclear quantum dynamics. 
However, adiabatic electronic states obtained from electronic structure computations necessarily involve random phases\cite{zhou2020a}, or more generally, random gauge fixings.  The indeterminacy of electronic wavefunction phases hamper the nonadiabatic wavepacket dynamics modeling, and have to be corrected either for constructing a quasi-diabatic model or for performing nuclear wavepacket dynamics. This is because the randomness in the phase will be transferred to any quantity defined using the many-electron wavefunction such as the nonadiabatic coupling and the transition dipole moment.  
This random phase can occur in every call of the matrix diagonalization subroutines, and thus,  cannot be avoided in ab initio modeling. 


A further complication arises from the presence of CIs, i.e., electronic degeneracy. In the presence of a CI, even when the time-reversal symmetry is preserved, it is impossible to find a global smooth gauge for real electronic wavefunctions. If such a global smooth gauge can be found, it would lead to a zero nonadiabatic coupling, and thus the geometric phase would be 0. This contradicts a geometric phase of $\pi$ encircling around a CI. it is then necessarily to add a nuclear-dependent phase to make the electronic wave functions globally smooth except for the CI seam.  

The problem of random phases becomes even more challenging when the electronic wave function is complex-valued. This occurs when the Hamiltonian does not have time-reversal symmetry, then the random phases becomes a complex phase instead of a sign.  Despite previous attempts, there is yet a straightforward way to fix the random phases for multi-state high-dimensional systems and for complex electronic wavefunctions \cite{akimov2018, lindefelt2004}.

Here we address the random phase problem in modeling conical intersection dynamics by integrating the local diabatic representation (LDR) and electronic structure computations.
The LDR was recently proposed to remove singular derivative couplings, both first- and second-order, from nonadiabatic conical intersection dynamics \cite{gu2023b}.  It has been demonstrated to provide numerically exact results for vibronic models \cite{gu2023b}.
We show that in the ab initio LDR, it is possible to perform exact modeling of conical intersection dynamics even using the adiabatic electronic states with phases randomly assigned during the electronic structure computations. 
This is demonstrated by an exact calculation of the two-dimensional Shin-Metiu model, a prototypical model for proton-coupled electron transfer \cite{shin1995}. 
Furthermore, it is shown that the LDR can provide exact results even when there is an external magnetic field where the electronic wavefunction becomes complex-valued and the random phase problem becomes more severe.   
Our results provide a simple approach to integrating the electronic structure computations into the conical intersection dynamics, thus paving the way for ab initio modeling of conical intersection dynamics.

Atomic units are used throughout $\hbar = e = m_\text{e} = 1$. 

\section{Theory and Method}
Ab initio conical intersection dynamics aims to solve the  \tdse for an entire molecule 
\be
i \frac{\partial}{\partial t} \ket{\Psi(t)} = H \ket{\Psi(t)}
\ee
where $\ket{\Psi(t)}$ is the electron-nuclear (vibronic) state. 
The molecular Coulomb Hamiltonian
\be
   H = T_\text{e} + T_\text{N} + {V}_\text{ee} + {V}_\text{eN} + {V}_\text{NN}, 
    \ee
contains the electronic and nuclear kinetic energy operators $T_\text{e}$ and $T_\text{N}$,  represents the repulsive electron-electron interaction $V_\text{ee}$,  the electron-nuclear Coulomb attraction  $V_\text{eN}$, and nuclear repulsion $V_\text{NN}$.

In the adiabatic representation, the molecular wavefunction is given by the Born-Huang expansion \cite{tannor2007}, 
\be 
\Psi(\bf r, \bf R, t) = \sum_\alpha \phi_\alpha(\bf r; \bf R) \chi_\alpha(\bf R, t)
\label{eq:bhe}
\ee 
where $\alpha$ runs over the adiabatic electronic states, usually truncated to a few low-lying excited states. 
In the Born-Huang ansatz, the adiabatic electronic states are required to be continuous with respect to the nuclear configurations. This is because when inserted back into the time-dependent molecular \se, the nuclear kinetic energy operator will act on the adiabatic states. 
However, in ab inito modeling, the adiabatic states are not smooth. 
The sign of the wavefunction is randomly determined during the electronic structure calculations.



\subsection{Unavoidable Random phases}

The random phases in the adiabatic electronic states arise from the gauge freedom. When the time-reversal symmetry is satisfied
\be \mc{T}H_\text{e} \mc{T}^{-1} = H_\text{e}
\ee
where $\mc{T}$ is the time-reversal operator, the electronic eigenstates can always be chosen to be real. The time-reversal operator is simply the complex conjugation $\mc{K}$ for spinless particles \cite{wang2018b}. However, even when the adiabatic states are constrained to be real, they cannot be uniquely determined. If $\psi_\alpha(\bf r; \bf R)$ is an eigenstate of the electronic Hamiltonian at nuclear geometry $H_\text{e}(\bf R)$, then $-\psi_{\alpha}(\bf r)$ is also an eigenstate. Therefore, there is a $\mathbb{Z}_2 = \set{1,-1}$ gauge freedom.  This implies that the adiabatic electronic states obtained from electronic structure calculations are non-smooth, and thus cannot be used in the Born-Huang expansion, which requires a smooth gauge. 

When the electronic Hamiltonian does not satisfy the time-reversal symmetry (e.g., in the presence of a magnetic field), the electronic states are complex. In this case, the gauge freedom becomes U(1); that is, if $\psi(\bf r)$ is an eigenstate, then $e^{i \theta} \psi(\bf r)$ is also an eigenstate for any $\theta\in [0, 2\pi)$. The adiabatic states in the presence of a magnetic field cannot be uniquely determined up to a local phase. Such states cannot be directly used in the adiabatic representation.

\subsection{Local Diabatic Representation}
We now show that in the LDR, it is possible to directly use the nonsmooth adiabatic electronic states obtained in electronic structure computations. 
We first use discrete variable representations (DVR) for the nuclear coordinate operators $\bf R$ \cite{littlejohn2002, light2000}. 
For each nuclear degree of freedom $q$, a DVR consists of a finite set, of basis functions $\set{\chi_n(q)}_{n=1}^{N}$ and a set of grid points $\set{q_i}$.  



The main idea of the LDR is to use the grid points defined in the DVR as reference nuclear geometries to define adiabatic electronic basis states, i.e., 
\be
H_\text{e}( \bf R_{\bf n})\ket{\phi_\alpha( \bf R_{\bf n})} = V_\alpha(\bf R_{\bf n}) \ket{\phi_\alpha( \bf R_{\bf n})} 
\ee
where $H_\text{e} = H - T_\text{N}$ is the electronic Hamiltonian, the full molecular Hamiltonian subtracting the nuclear kinetic energy operator, 
and $V_\alpha(\bf R)$ is  the $\alpha$th adiabatic potential energy surface.

The  LDR ansatz for the full molecular wavefunction is now given by
\be
\begin{split}
\Psi(\bf r, \bf R, t) &= \sum_{\bf n} {\sum_\alpha C_{\bf n\alpha}(t) \phi_\alpha(\bf r; \bf R_{\bf n})} \chi_{\bf n}(\bf R) \\ 
&\equiv \sum_{\bf n, \alpha} C_{\bf n\alpha}(t) \braket{\bf r, \bf R | \bf n \alpha}
\end{split}
\label{eq:ansatz}
\ee
where we have introduced a shorthand notation $\ket{\bf n \alpha} \equiv \phi_\alpha(\bf r; \bf R_{\bf n}) \chi_{\bf n}(\bf R)$.
In contrast to the adiabatic representation in \cref{eq:bhe}, in the LDR ansatz, we do not impose any smooth conditions on $\phi_\alpha(\bf R_{\bf n})$. Here it is understood that the adiabatic electronic states are directly coming from the electronic structure solver with random phases.

The formal solution of the time-dependent molecular \text{\se} is given by
\be 
\ket{\Psi(t)} = e^{-i H t} \ket{\Psi_0}. 
\label{eq:111}
\ee 
where $\ket{\Psi_0}$ is the initial vibronic state. 
To solve \cref{eq:111}, we apply the Strang splitting to the full short-time molecular propagator \cite{gu2024a}, 
\be
e^{-i H \Delta t} \approx  e^{- i H_\text{e}(\bf R) \Delta t/2}   e^{- i T_\text{N} \Delta t} e^{- i H_\text{e}(\bf R) \Delta t/2}
\label{eq:ldr}
\ee 
The propagator with the electronic Hamiltonian $H_\text{e}$ can be computed as  
\be
 e^{-i H_\text{e}(\bf R) \Delta t} \ket{\Psi(t)} \approx 
\sum_{\bf n, \alpha} C_{\bf n\alpha}(t) e^{-i V_{\alpha}(\bf R_{\bf n}) \Delta t } \ket{\bf n\alpha}
\ee 
where we have made use of a diagonal approximation 
\be 
\braket{\bf m \beta | H_\text{e}(\bf R) | \bf n \alpha} \approx V_\alpha(\bf R_{\bf n}) \delta_{\beta \alpha}  \delta_{\bf m \bf n}
\ee 
For the nuclear kinetic energy propagator, 
\be 
\braket{ \bf{m}\beta| e^{- i T_\text{N}\Delta t} | \alpha \bf{n}}   
= {A}{_{\bf m \beta, \bf n \alpha}}\braket{\chi_\bf{m} | e^{-i T_\text{N} \Delta t}| \chi_\bf{n}}_{\bf R} 
\ee 
The electronic overlap matrix, encoding all nonadiabatic and geometric phase effects, is defined as  
\be
A_{\bf{m}\beta, \bf{n}\alpha} = \braket{\phi_{\beta}(\bf R_\bf{m}) | \phi_\alpha(\bf R_\bf{n}) }_{\bf r},
\label{eq:overlap}
\ee
where $\braket{\cdots}_{\bf r}$ ($\braket{\cdots}_{\bf R}$) denotes the integration over electronic (nuclear) degrees of freedom.  

Here the matrix elements of the exponential kinetic energy operator $\sbr{e^{-i T_\text{N} \Delta t}}_\bf{m n} $ can be analytically computed in the finite basis and transformed to the DVR basis set. The split-operator method in the LDR differs from the the conventional one for adiabatic dynamics\cite{kosloff1988} in that the fast Fourier transform cannot be used here due to the presence of electronic overlap matrix. 

While LDR also employs adiabatic electronic states as the electronic basis, 
it is immune to the random phases in the adiabatic wavefunctions because the propagation of LDR  does not involve nuclear derivative terms. The phase will only affects the electronic overlap matrix, which will then be compensated by the expansion coefficients $C_{\bf n\alpha}$.

\section{Model and Computations}

To demonstrate that LDR can resolve the random phase problem, we solve the two-dimensional (2D) Shin-Meiu model numerically exactly by the LDR method.
The extended 2D Shin-Metiu model is a prototypical model for proton-coupled electron transfer dynamics. \cite{shin1995, shin1996, min2014}
This model, depicted in Figure 1a, consists of three protons and one electron; Two protons are fixed at positions ($\pm{L}/2, 0$), while the third proton and the electron are free to move in 2D space. 
The variables $\bf{R}=(X,Y)$ and $\bf{r}=(x,y)$ represent the positions of the movable proton and the electron, respectively. The full Hamiltonian of the system is given by
\begin{align}
    {H}(\bf{r},\bf{R}) = \frac{\bf p^2}{2} + \frac{\bf P^2}{2M} + V(\bf r,\bf R)
\end{align}
where $\bf p=-i \frac{\partial}{\partial \bf r}$ and $\bf P=-i  \frac{\partial}{\partial \bf R} $ are the momentum operators for the electron and the moving proton with mass $M$. The potential energy ${V}(\bf{r},\bf{R})$ reads 
\be
{V}(\bf{r},\bf{R}) =V_\text{eN} +  V_\text{NN} + \left(\frac{R}{R_0}\right)^4,
\label{eq:sm}
\ee
Soft Coulomb potentials are used for the electron-nucleus interaction potential, $V_\text{en}(\bf r, \bf R) = -1/\sqrt{a+(\bf r - \bf R)^2}$, and the nucleus-nucleus interaction potential, $V_\text{nn}(\bf R, \bf R_i) = 1/\sqrt{b+(\bf R - \bf R_i)^2}$,  $\bf R_i, i =1 ,2 $ refers to the location of the two fixed ions, and parameters $a=0.5$ and $b=10.0$.
The last quartic potential in \cref{eq:sm} guarantees that the system remains bound, with $R_0 = 3.5$ a.u.

The Born-Oppenheimer adiabatic potential energy surfaces $V_{\alpha}(\bf{R})$ are defined as eigenvalues of the electronic \se:
\be
\left[\frac{\bf {p}^2}{2}+V(\bf{r};\bf{R})\right]\phi_\alpha(\bf{r};\bf{R}) = V_{\alpha}(\bf R) \phi_\alpha(\bf{r};\bf{R}),
\ee
where $\phi_\alpha(\bf r; \bf R)$ are the electronic eigenfunctions. 
The electronic structure problem for the single electron for each nuclear configuration is computed in 2D uniform grids. 
To achieve convergence, 127 grid points in the range of [-6, 6] a.u. are used for the electronic degrees of freedom.
The first and second electronic eigenstates at $\bf R = (0.0,1.21875)$ are shown in Figure S1a.
By scanning nuclear coordinates $\bf R$,  the first and second excited adiabatic potential energy surfaces (APESs) are constructed.
The APESs are shown in Figure 1b, with the energy at $\bf{R}_{CI}$ being -7.6921 eV.
Due to the identical interaction potentials among the three protons, the CI position is \(\mathbf{R}_{CI}^{\pm} = (0, \pm Y_{CI})\), with \(Y_{CI} = \sqrt{3}/2L = 1.2\) a.u.


\begin{figure}[h] 
    \centering
    \includegraphics[width=0.45\textwidth, trim=0 0 0 10, clip]{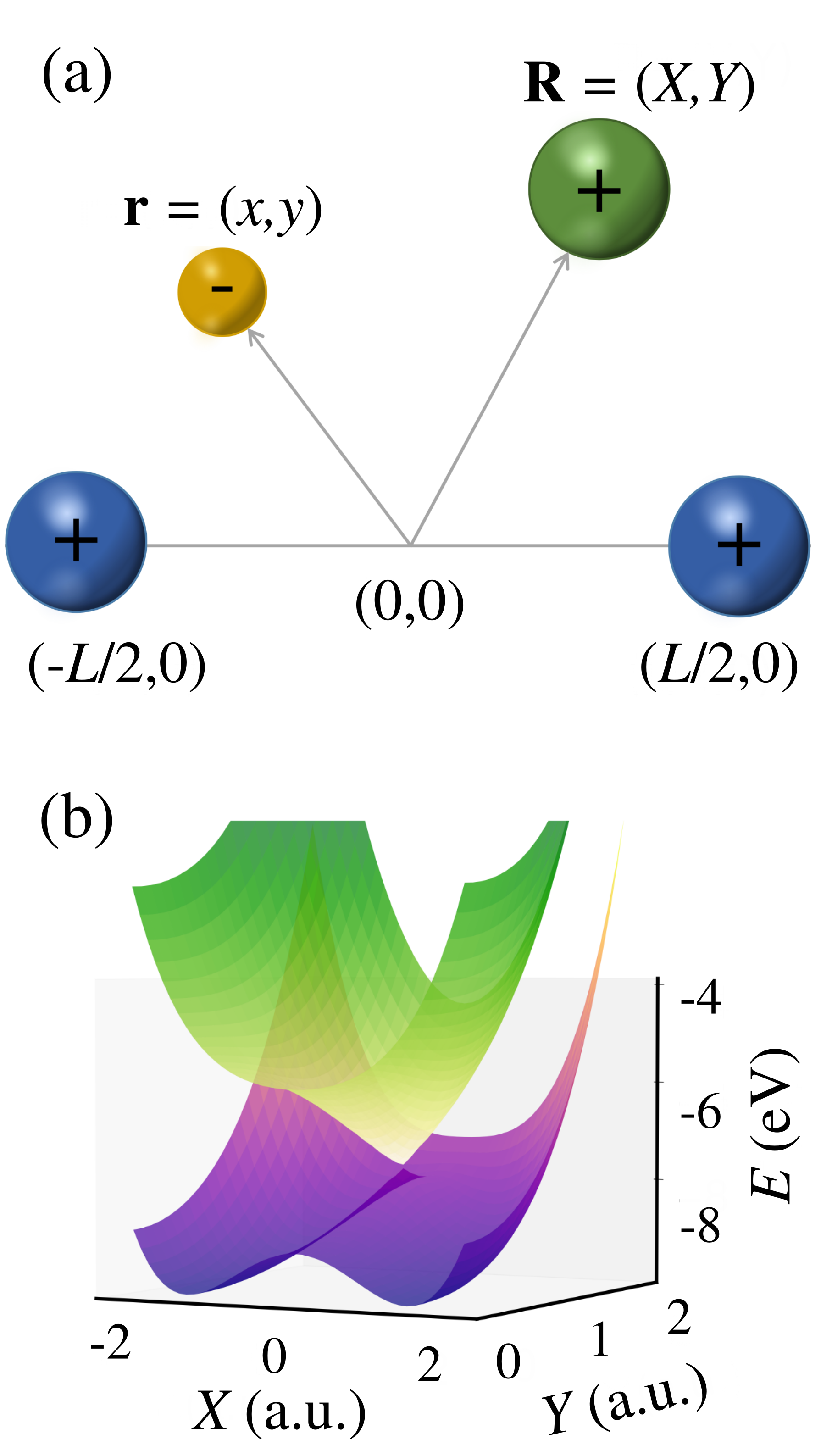} 
    \caption{(a) Schematic of the two-dimensional Shin-Metiu model ($L=4\sqrt{3}/5$ a.u.) and (b) adiabatic potential energy surfaces for the first and second excited states.}
\end{figure}

By scanning the 2D nuclear space, the adiabatic electronic states and electron overlap matrix are obtained 
which are subsequently inserted to the LDR method (\cref{eq:ldr}) to solve for the exact conical intersection dynamics.
63 grid points in the range of [-3, 3] a.u. are used for each nuclear degrees of freedom.
The electronic overlap submatrix $A_{m_1 m_2 \beta, n_1 n_2 \alpha}$ for $m_2= 0, \beta =2, n_2= 0, \alpha =2 $ is shown in Figure 2a, 
which clearly shows the random signs on the adiabatic states.

We use the following initial state 
$\Psi_0(\bf r;\bf R) = \phi_{2}(\bf{r;R_0}) \chi(\bf{R}-\bf{R_0})$
where, $\phi_2(\bf {r;R_0})$ represents the second electronic eigenstate at the nuclear configuration $\bf{R_0}=(0.0,1.78125)$, and
$\chi(\bf{R}-\bf{R_0}) = \sqrt{\pi/\gamma} e^{-\frac{\gamma}{2}(\bf R -\bf R_0)^2} $ 
is a Gaussian nuclear wavepacket centered at $\bf{R_0}$ with width $\gamma = 18$ a.u.

The initial state is given in the crude adiabatic representation. It can be transformed to the LDR by  
\begin{align} 
\ket{\Psi_0} = \sum_{\bf n, \beta} \chi_\bf{n} A_{\bf{n}\beta , \bf{n_0} \alpha } \ket{ \bf{n} \beta}  
\end{align}
The overlap matrix element $A_{\bf{n} \beta , \bf{n_0} \alpha}$ accounts for the phase relationship between the electronic states at different nuclear configurations. 

Here, both $A_{\bf{n} \beta , \bf{n_0} \alpha}$ and $\ket{\bf{n} \beta }$ contain the random phases, and their product cancels out the effect of random phase leading to a smooth molecular wavefunction. The initial wavefunctions are starting in the second excited state. This configuration positions the initial wavepacket centrally between the two fixed ions ($X_0=0.0$) with a displacement of $Y_0$ along the $Y$-axis, located above the CI. 

\begin{figure*}[h] 
    \centering
    \includegraphics[width=0.95\textwidth, trim=0 55 0 35, clip]{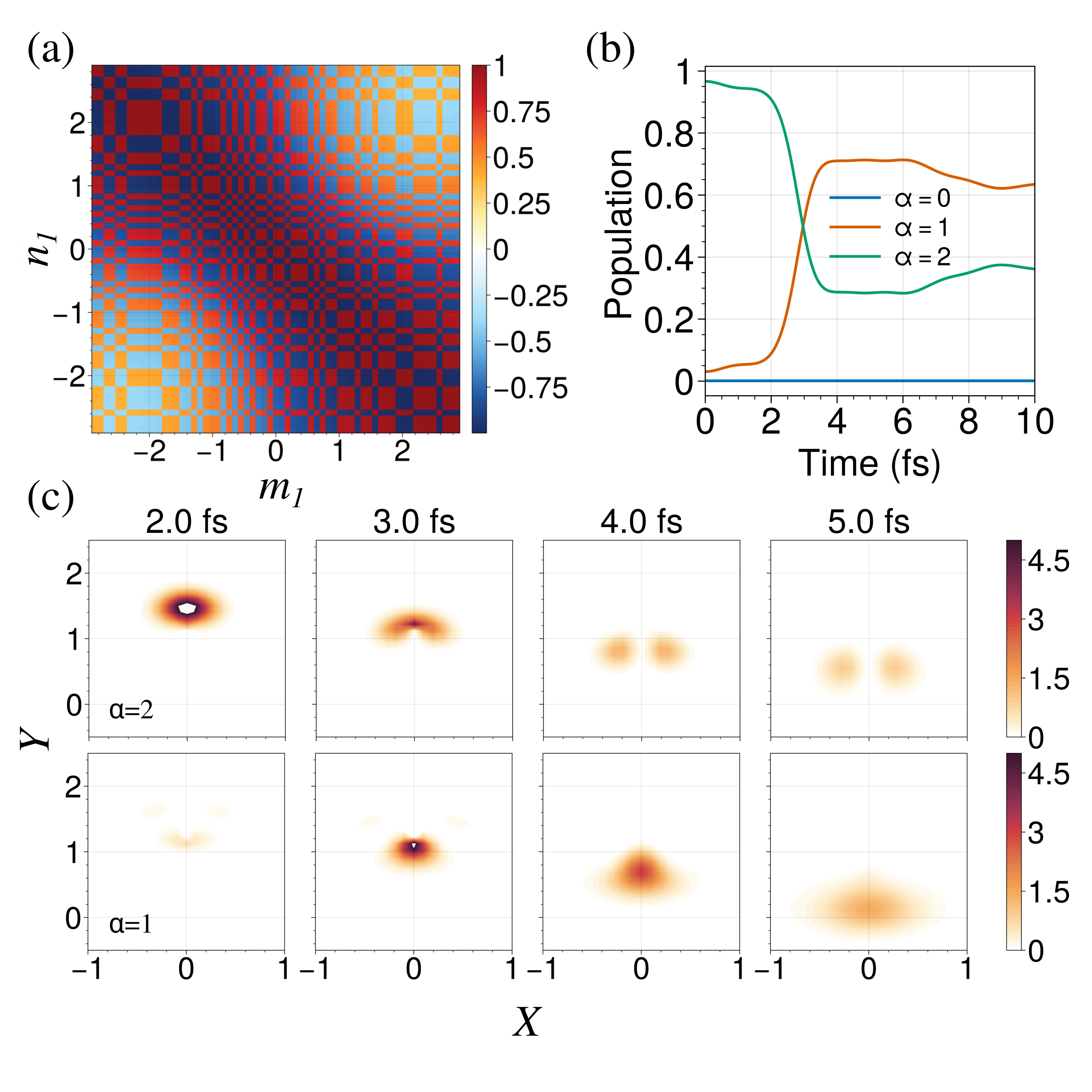} 
    \caption{(a) The electronic overlap submatrix $A_{m_1 m_2 \beta, n_1 n_2 \alpha}$ for $m_2= 0, \beta =2, n_2= 0, \alpha =2 $ in field-free case,(b) population dynamics of the ground state and excited states within 10 fs, and (c) the wavepacket dynamics $|\chi_{\bf{n} \alpha}|^2$ of excited states ($\alpha = 1,2$) through the CI. }
\end{figure*}

The adiabatic electronic states are directly inserted to the LDR method  to perform nonadiabatic wavepacket dynamics around CIs.
Figure 2b displays the adiabatic electronic populations dynamics in the two excited electronic states. At $\sim$ 2 fs, the nuclear wavepacket encounters the CI at (0, 1.2) a.u., making nonadiabatic transition to the first excited state, 
resulting in nonadiabatic relaxation. This transfer happens extremely rapidly, completing in approximately 4 fs.
Figure 2c depicts  the nuclear wavepacket dynamics on the adiabatic electronic states ($\alpha =1, 2$). 
Within the first 4 fs, the proton moves downwards on the second excited state, encounters the CI, and transits to the first excited state. The geometric phase manifest as the nodal line along the $X = 0$. 
These results are in excellent agreement with the results in Ref. \citenum{hader2017}, demonstrating that the LDR method have successfully addressed the random phase problem.

\subsection{Magnetic Field}

We now apply a static magnetic field to the 2D Shin-Metiu model to demonstate the capability of our method in accurately modeling conical intersection dynamics even when the adiabatic electronic states are complex.
In the presence of a magnetic field $\bf B = (0, 0, B) = B \hat{\bf z}$, where $\hat{\bf z}$ is the unit vector in the $z$-axis, the vector potential in the Landau gauge is given by 
\be 
\bf A = (0, Bx, 0)^\text{T}
\ee 
where superscript T denotes transpose. Alternatively, one can also choose the symmetry gauge where $\bf A = \half \bf B \times \bf r = \half (-By, Bx, 0)$. 

The electronic Hamiltonian in the presence of magnetic field is given by
\be 
\begin{split}
H_\text{e} &= \frac{p_x^2}{2} + \frac{1}{2} \del{p_y + Bx}^2 + V(\bf r) \\ 
&= \frac{\bf p^2}{2} + V(\bf r) + \half B^2 x^2 + Bx p_y 
\end{split}
\ee 

A crucial difference due to the magnetic field is that the time-reversal symmetry is broken such that electronic states becomes complex-valued. Figure S1b shows the first and second electronic eigenstates at $\bf R = (0.0,1.21875)$  with a magnetic field $B = 40$ T. The adiabatic electronic states now carry a random phase $e^{i \theta_{\bf n}}$.   
In this scenario, the gauge freedom becomes more complicated than field-free case.
The phase of the complex electronic overlap matrix $A$ are shown in Figure 3a, which clearly shows the random of the phase.
\begin{figure*}[h] 
    \centering
    \includegraphics[width=0.95\textwidth, trim=0 30 0 20, clip]{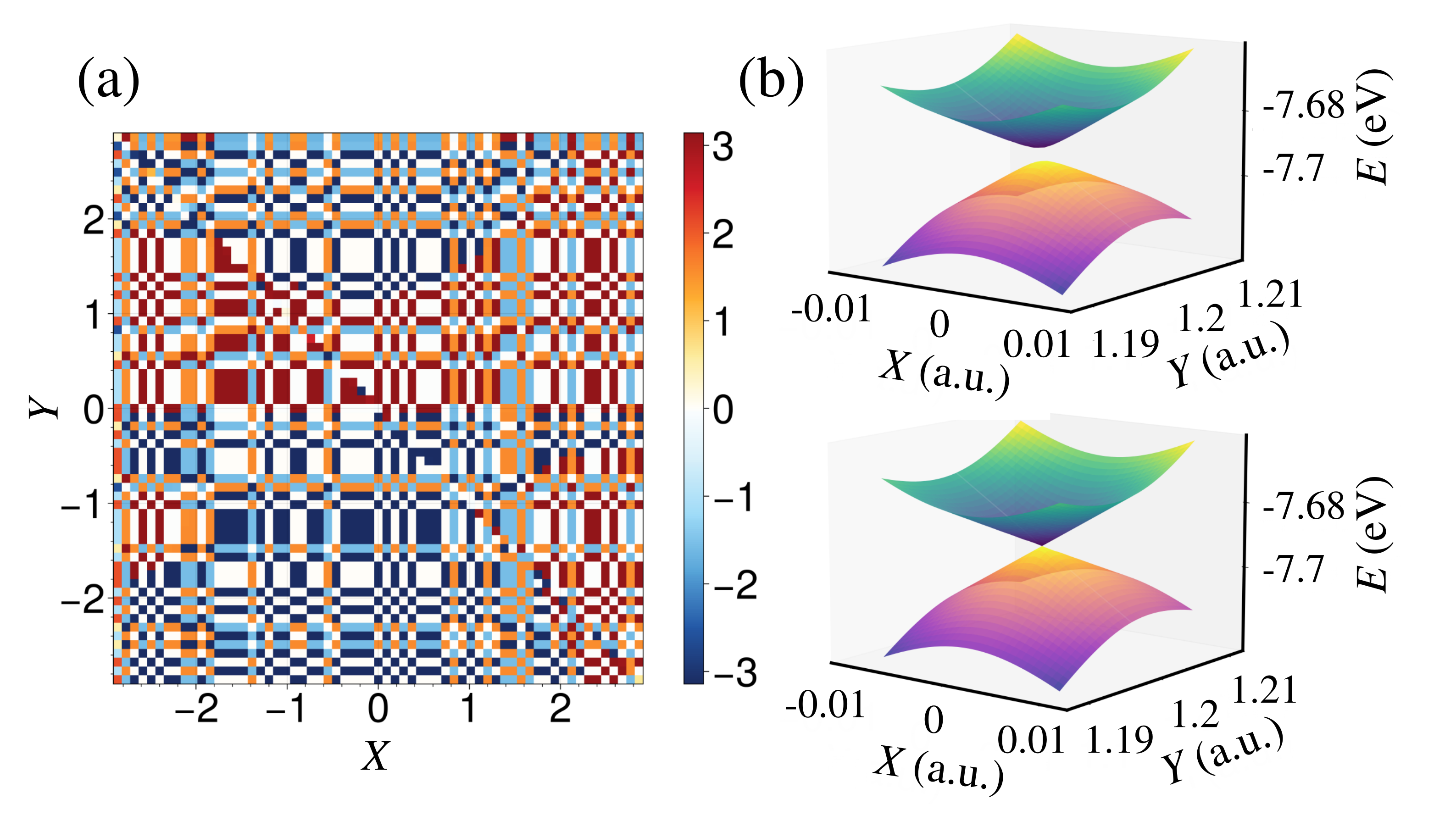} 
    \caption{(a) The random phase of electronic overlap matrix in the presence of a magnetic field, 
    and (b) APESs around CI with (top panel) and without a magnetic field (bottom panel). The magnetic field opens a gap at the pristine CI.}
\end{figure*}

To solve the nonadiabatic CI dynamics, we follow similar steps as the field-free case. The APESs in the presence of the magnetic field is firstly constructed by diagonalizating the electronic Hamiltonian in a DVR basis set.   The first and second excited APESs are shown in Figure 3b.  A small region centered around the pristine CI   is chosen  to highlight the magnetic field effects. 
As shown, the magnetic field opens an energy gap at the CI, turning the original electronic degeneracy   to near-degeneracy.


After constructing the APESs and electronic overlap matrix, the LDR method is used to model the non-adiabatic dynamics in the presence of magnetic field, using the same nuclear grids as in the field-free case. The adiabatic population and nuclear wavepacket dynamics of the first and second excited states are shown in Figure S2, which closely resemble those in the field-free case. This implies that although the CI topology is altered, the magnetic field effect is too weak to impact the overall nonadiabatic dynamics. Note that the geometric phase effects remains intact despite the disappearance of an exact intersection. 
This indicates that the LDR method can correctly handle the case where the electronic states are complex-valued with completely random phases.




\section{Summary}

To summarize, by an exact modeling of the two-dimensional Shin-Metiu model, we have shown that the ab initio local adiabatic representation (LDR) provides a simple solution to the random gauge problem in modeling conical intersection dynamics.  
We further show that even in the presence of a magnetic field, where the electronic states are complex-valued, our method still provides an exact result.

Our results clearly show that the LDR is a straightforward and efficient approach for simulating the nonadiabatic conical intersection dynamics in random gauges, regardless of whether gauge freedom is $\mathbb{Z}_2$ or U(1). 
By resolving the random phase problem, our method provides a straightforward approach to integrating electronic structure simulation with conical intersection dynamics.

\begin{acknowledgement}
	This work is supported by the National Natural Science Foundation of China (Grant No. 92356310).
\end{acknowledgement}

\bibliography{electronic_structure, optics,qchem,dynamics,cavity}

\providecommand{\latin}[1]{#1}
\makeatletter
\providecommand{\doi}
  {\begingroup\let\do\@makeother\dospecials
  \catcode`\{=1 \catcode`\}=2 \doi@aux}
\providecommand{\doi@aux}[1]{\endgroup\texttt{#1}}
\makeatother
\providecommand*\mcitethebibliography{\thebibliography}
\csname @ifundefined\endcsname{endmcitethebibliography}
  {\let\endmcitethebibliography\endthebibliography}{}
\begin{mcitethebibliography}{27}
\providecommand*\natexlab[1]{#1}
\providecommand*\mciteSetBstSublistMode[1]{}
\providecommand*\mciteSetBstMaxWidthForm[2]{}
\providecommand*\mciteBstWouldAddEndPuncttrue
  {\def\EndOfBibitem{\unskip.}}
\providecommand*\mciteBstWouldAddEndPunctfalse
  {\let\EndOfBibitem\relax}
\providecommand*\mciteSetBstMidEndSepPunct[3]{}
\providecommand*\mciteSetBstSublistLabelBeginEnd[3]{}
\providecommand*\EndOfBibitem{}
\mciteSetBstSublistMode{f}
\mciteSetBstMaxWidthForm{subitem}{(\alph{mcitesubitemcount})}
\mciteSetBstSublistLabelBeginEnd
  {\mcitemaxwidthsubitemform\space}
  {\relax}
  {\relax}

\bibitem[Truhlar and Mead(2003)Truhlar, and Mead]{truhlar2003}
Truhlar,~D.~G.; Mead,~C.~A. Relative Likelihood of Encountering Conical
  Intersections and Avoided Intersections on the Potential Energy Surfaces of
  Polyatomic Molecules. \emph{Phys. Rev. A} \textbf{2003}, \emph{68},
  032501\relax
\mciteBstWouldAddEndPuncttrue
\mciteSetBstMidEndSepPunct{\mcitedefaultmidpunct}
{\mcitedefaultendpunct}{\mcitedefaultseppunct}\relax
\EndOfBibitem
\bibitem[Domcke and Yarkony(2012)Domcke, and Yarkony]{domcke2012}
Domcke,~W.; Yarkony,~D.~R. Role of {{Conical Intersections}} in {{Molecular
  Spectroscopy}} and {{Photoinduced Chemical Dynamics}}. \emph{Annu. Rev. Phys.
  Chem.} \textbf{2012}, \emph{63}, 325--352\relax
\mciteBstWouldAddEndPuncttrue
\mciteSetBstMidEndSepPunct{\mcitedefaultmidpunct}
{\mcitedefaultendpunct}{\mcitedefaultseppunct}\relax
\EndOfBibitem
\bibitem[K{\"o}ppel \latin{et~al.}(2004)K{\"o}ppel, Domcke, and
  Cederbaum]{koppel2004}
K{\"o}ppel,~H.; Domcke,~W.; Cederbaum,~L.~S. \emph{Advanced {{Series}} in
  {{Physical Chemistry}}}; WORLD SCIENTIFIC, 2004; Vol.~15; pp 323--367\relax
\mciteBstWouldAddEndPuncttrue
\mciteSetBstMidEndSepPunct{\mcitedefaultmidpunct}
{\mcitedefaultendpunct}{\mcitedefaultseppunct}\relax
\EndOfBibitem
\bibitem[Larson \latin{et~al.}(2020)Larson, Sj{\"o}qvist, and
  {\"O}hberg]{larson2020}
Larson,~J.; Sj{\"o}qvist,~E.; {\"O}hberg,~P. \emph{Conical {{Intersections}} in
  {{Physics}}: {{An Introduction}} to {{Synthetic Gauge Theories}}}; Lecture
  {{Notes}} in {{Physics}}; Springer International Publishing: Cham, 2020; Vol.
  965\relax
\mciteBstWouldAddEndPuncttrue
\mciteSetBstMidEndSepPunct{\mcitedefaultmidpunct}
{\mcitedefaultendpunct}{\mcitedefaultseppunct}\relax
\EndOfBibitem
\bibitem[Levine and Mart{\'i}nez(2007)Levine, and Mart{\'i}nez]{levine2007}
Levine,~B.~G.; Mart{\'i}nez,~T.~J. Isomerization {{Through Conical
  Intersections}}. \emph{Annu. Rev. Phys. Chem.} \textbf{2007}, \emph{58},
  613--634\relax
\mciteBstWouldAddEndPuncttrue
\mciteSetBstMidEndSepPunct{\mcitedefaultmidpunct}
{\mcitedefaultendpunct}{\mcitedefaultseppunct}\relax
\EndOfBibitem
\bibitem[Mead and Truhlar(1979)Mead, and Truhlar]{mead1979}
Mead,~C.~A.; Truhlar,~D.~G. On the Determination of {{Born}}--{{Oppenheimer}}
  Nuclear Motion Wave Functions Including Complications Due to Conical
  Intersections and Identical Nuclei. \emph{The Journal of Chemical Physics}
  \textbf{1979}, \emph{70}, 2284--2296\relax
\mciteBstWouldAddEndPuncttrue
\mciteSetBstMidEndSepPunct{\mcitedefaultmidpunct}
{\mcitedefaultendpunct}{\mcitedefaultseppunct}\relax
\EndOfBibitem
\bibitem[Mead(1992)]{mead1992}
Mead,~C.~A. The Geometric Phase in Molecular Systems. \emph{Rev. Mod. Phys.}
  \textbf{1992}, \emph{64}, 51--85\relax
\mciteBstWouldAddEndPuncttrue
\mciteSetBstMidEndSepPunct{\mcitedefaultmidpunct}
{\mcitedefaultendpunct}{\mcitedefaultseppunct}\relax
\EndOfBibitem
\bibitem[{Joubert-Doriol} \latin{et~al.}(2013){Joubert-Doriol}, Ryabinkin, and
  Izmaylov]{joubert-doriol2013}
{Joubert-Doriol},~L.; Ryabinkin,~I.~G.; Izmaylov,~A.~F. Geometric Phase Effects
  in Low-Energy Dynamics near Conical Intersections: {{A}} Study of the
  Multidimensional Linear Vibronic Coupling Model. \emph{J. Chem. Phys.}
  \textbf{2013}, \emph{139}, 234103\relax
\mciteBstWouldAddEndPuncttrue
\mciteSetBstMidEndSepPunct{\mcitedefaultmidpunct}
{\mcitedefaultendpunct}{\mcitedefaultseppunct}\relax
\EndOfBibitem
\bibitem[Bersuker(2006)]{bersuker2006}
Bersuker,~I. \emph{The {{Jahn-Teller Effect}}}; Cambridge University Press:
  Cambridge, 2006\relax
\mciteBstWouldAddEndPuncttrue
\mciteSetBstMidEndSepPunct{\mcitedefaultmidpunct}
{\mcitedefaultendpunct}{\mcitedefaultseppunct}\relax
\EndOfBibitem
\bibitem[Gu and Mukamel(2020)Gu, and Mukamel]{gu2020b}
Gu,~B.; Mukamel,~S. Cooperative {{Conical Intersection Dynamics}} of {{Two
  Pyrazine Molecules}} in an {{Optical Cavity}}. \emph{J. Phys. Chem. Lett.}
  \textbf{2020}, \emph{11}, 5555--5562\relax
\mciteBstWouldAddEndPuncttrue
\mciteSetBstMidEndSepPunct{\mcitedefaultmidpunct}
{\mcitedefaultendpunct}{\mcitedefaultseppunct}\relax
\EndOfBibitem
\bibitem[Gu and Mukamel(2020)Gu, and Mukamel]{gu2020c}
Gu,~B.; Mukamel,~S. Manipulating Nonadiabatic Conical Intersection Dynamics by
  Optical Cavities. \emph{Chem. Sci.} \textbf{2020}, \emph{11},
  1290--1298\relax
\mciteBstWouldAddEndPuncttrue
\mciteSetBstMidEndSepPunct{\mcitedefaultmidpunct}
{\mcitedefaultendpunct}{\mcitedefaultseppunct}\relax
\EndOfBibitem
\bibitem[Xie \latin{et~al.}(2019)Xie, Malbon, Guo, and Yarkony]{xie2019}
Xie,~C.; Malbon,~C.~L.; Guo,~H.; Yarkony,~D.~R. Up to a {{Sign}}. {{The
  Insidious Effects}} of {{Energetically Inaccessible Conical Intersections}}
  on {{Unimolecular Reactions}}. \emph{Acc. Chem. Res.} \textbf{2019},
  \emph{52}, 501--509\relax
\mciteBstWouldAddEndPuncttrue
\mciteSetBstMidEndSepPunct{\mcitedefaultmidpunct}
{\mcitedefaultendpunct}{\mcitedefaultseppunct}\relax
\EndOfBibitem
\bibitem[Zhou \latin{et~al.}(2020)Zhou, Jin, Qiu, Rappe, and
  Subotnik]{zhou2020a}
Zhou,~Z.; Jin,~Z.; Qiu,~T.; Rappe,~A.~M.; Subotnik,~J.~E. Phase Problem: {{A
  Robust}} and {{Unified Solution}} for {{Choosing}} the {{Phases}} of
  {{Adiabatic States}} as a {{Function}} of {{Geometry}}: {{Extending Parallel
  Transport Concepts}} to the {{Cases}} of {{Trivial}} and {{Near-Trivial
  Crossings}}. \emph{J. Chem. Theory Comput.} \textbf{2020}, \emph{16},
  835--846\relax
\mciteBstWouldAddEndPuncttrue
\mciteSetBstMidEndSepPunct{\mcitedefaultmidpunct}
{\mcitedefaultendpunct}{\mcitedefaultseppunct}\relax
\EndOfBibitem
\bibitem[Akimov(2018)]{akimov2018}
Akimov,~A.~V. A {{Simple Phase Correction Makes}} a {{Big Difference}} in
  {{Nonadiabatic Molecular Dynamics}}. \emph{J. Phys. Chem. Lett.}
  \textbf{2018}, \emph{9}, 6096--6102\relax
\mciteBstWouldAddEndPuncttrue
\mciteSetBstMidEndSepPunct{\mcitedefaultmidpunct}
{\mcitedefaultendpunct}{\mcitedefaultseppunct}\relax
\EndOfBibitem
\bibitem[Lindefelt \latin{et~al.}(2004)Lindefelt, Nilsson, and
  Hjelm]{lindefelt2004}
Lindefelt,~U.; Nilsson,~H.-E.; Hjelm,~M. Choice of Wavefunction Phases in the
  Equations for Electric-Field-Induced Interband Transitions. \emph{Semicond.
  Sci. Technol.} \textbf{2004}, \emph{19}, 1061--1066\relax
\mciteBstWouldAddEndPuncttrue
\mciteSetBstMidEndSepPunct{\mcitedefaultmidpunct}
{\mcitedefaultendpunct}{\mcitedefaultseppunct}\relax
\EndOfBibitem
\bibitem[Gu(2023)]{gu2023b}
Gu,~B. A {{Discrete-Variable Local Diabatic Representation}} of {{Conical
  Intersection Dynamics}}. \emph{J. Chem. Theory Comput.} \textbf{2023},
  \emph{19}, 6557--6563\relax
\mciteBstWouldAddEndPuncttrue
\mciteSetBstMidEndSepPunct{\mcitedefaultmidpunct}
{\mcitedefaultendpunct}{\mcitedefaultseppunct}\relax
\EndOfBibitem
\bibitem[Shin and Metiu(1995)Shin, and Metiu]{shin1995}
Shin,~S.; Metiu,~H. Nonadiabatic Effects on the Charge Transfer Rate Constant:
  {{A}} Numerical Study of a Simple Model System. \emph{J. Chem. Phys.}
  \textbf{1995}, \emph{102}, 9285--9295\relax
\mciteBstWouldAddEndPuncttrue
\mciteSetBstMidEndSepPunct{\mcitedefaultmidpunct}
{\mcitedefaultendpunct}{\mcitedefaultseppunct}\relax
\EndOfBibitem
\bibitem[Tannor(2007)]{tannor2007}
Tannor,~D.~J. \emph{Introduction to {{Quantum Mechanics}}: {{A Time-Dependent
  Perspective}}}; University Science Books, 2007\relax
\mciteBstWouldAddEndPuncttrue
\mciteSetBstMidEndSepPunct{\mcitedefaultmidpunct}
{\mcitedefaultendpunct}{\mcitedefaultseppunct}\relax
\EndOfBibitem
\bibitem[Wang \latin{et~al.}(2018)Wang, Devereaux, and Chen]{wang2018b}
Wang,~Y.; Devereaux,~T.~P.; Chen,~C.-C. Theory of Time-Resolved {{Raman}}
  Scattering in Correlated Systems: {{Ultrafast}} Engineering of Spin Dynamics
  and Detection of Thermalization. \emph{Phys. Rev. B} \textbf{2018},
  \emph{98}, 245106\relax
\mciteBstWouldAddEndPuncttrue
\mciteSetBstMidEndSepPunct{\mcitedefaultmidpunct}
{\mcitedefaultendpunct}{\mcitedefaultseppunct}\relax
\EndOfBibitem
\bibitem[Littlejohn \latin{et~al.}(2002)Littlejohn, Cargo, Carrington,
  Mitchell, and Poirier]{littlejohn2002}
Littlejohn,~R.~G.; Cargo,~M.; Carrington,~T.; Mitchell,~K.~A.; Poirier,~B. A
  General Framework for Discrete Variable Representation Basis Sets. \emph{J.
  Chem. Phys.} \textbf{2002}, \emph{116}, 8691--8703\relax
\mciteBstWouldAddEndPuncttrue
\mciteSetBstMidEndSepPunct{\mcitedefaultmidpunct}
{\mcitedefaultendpunct}{\mcitedefaultseppunct}\relax
\EndOfBibitem
\bibitem[Light and Carrington~Jr.(2000)Light, and Carrington~Jr.]{light2000}
Light,~J.~C.; Carrington~Jr.,~T. \emph{Advances in {{Chemical Physics}}}; John
  Wiley \& Sons, Ltd, 2000; pp 263--310\relax
\mciteBstWouldAddEndPuncttrue
\mciteSetBstMidEndSepPunct{\mcitedefaultmidpunct}
{\mcitedefaultendpunct}{\mcitedefaultseppunct}\relax
\EndOfBibitem
\bibitem[Gu(2024)]{gu2024a}
Gu,~B. Generalized {{Optical Sum Rules}} for {{Light-Dressed Matter}}. \emph{J.
  Phys. Chem. Lett.} \textbf{2024}, 5580--5585\relax
\mciteBstWouldAddEndPuncttrue
\mciteSetBstMidEndSepPunct{\mcitedefaultmidpunct}
{\mcitedefaultendpunct}{\mcitedefaultseppunct}\relax
\EndOfBibitem
\bibitem[Kosloff(1988)]{kosloff1988}
Kosloff,~R. Time-Dependent Quantum-Mechanical Methods for Molecular Dynamics.
  \emph{J. Chem. Phys.} \textbf{1988}, \emph{92}, 2087--2100\relax
\mciteBstWouldAddEndPuncttrue
\mciteSetBstMidEndSepPunct{\mcitedefaultmidpunct}
{\mcitedefaultendpunct}{\mcitedefaultseppunct}\relax
\EndOfBibitem
\bibitem[Shin and Metiu(1996)Shin, and Metiu]{shin1996}
Shin,~S.; Metiu,~H. Multiple {{Time Scale Quantum Wavepacket Propagation}}:\,
  {{Electron}}-{{Nuclear Dynamics}}. \emph{J. Phys. Chem.} \textbf{1996},
  \emph{100}, 7867--7872\relax
\mciteBstWouldAddEndPuncttrue
\mciteSetBstMidEndSepPunct{\mcitedefaultmidpunct}
{\mcitedefaultendpunct}{\mcitedefaultseppunct}\relax
\EndOfBibitem
\bibitem[Min \latin{et~al.}(2014)Min, Abedi, Kim, and Gross]{min2014}
Min,~S.~K.; Abedi,~A.; Kim,~K.~S.; Gross,~E. K.~U. Is the {{Molecular Berry
  Phase}} an {{Artifact}} of the {{Born-Oppenheimer Approximation}}?
  \emph{Phys. Rev. Lett.} \textbf{2014}, \emph{113}, 263004\relax
\mciteBstWouldAddEndPuncttrue
\mciteSetBstMidEndSepPunct{\mcitedefaultmidpunct}
{\mcitedefaultendpunct}{\mcitedefaultseppunct}\relax
\EndOfBibitem
\bibitem[Hader \latin{et~al.}(2017)Hader, Albert, Gross, and Engel]{hader2017}
Hader,~K.; Albert,~J.; Gross,~E. K.~U.; Engel,~V. Electron-Nuclear Wave-Packet
  Dynamics through a Conical Intersection. \emph{J. Chem. Phys.} \textbf{2017},
  \emph{146}, 074304\relax
\mciteBstWouldAddEndPuncttrue
\mciteSetBstMidEndSepPunct{\mcitedefaultmidpunct}
{\mcitedefaultendpunct}{\mcitedefaultseppunct}\relax
\EndOfBibitem
\end{mcitethebibliography}

\end{document}